# Integrating Black Start Capabilities into Offshore Wind Farms by Grid-Forming Batteries

Daniela Pagnani*♦, Łukasz Kocewiak*, Jesper Hjerrild*, Frede Blaabjerg♦, Claus Leth Bak♦
*Electrical System Design and Grid Integration, Ørsted, Gentofte, Denmark
♦AAU Energy, Aalborg University, Aalborg, Denmark
{dapag, lukko, jeshj}@orsted.com,{fbl, clb}@energy.aau.dk

*Abstract*—Power systems are currently experiencing a transition towards decarbonisation through the large-scale deployment of renewable energy sources. These are gradually replacing conventional thermal power plants which today are the main providers of black start services. Consequently, in case of a total/partial blackout, conventional black-start resources may not be ready for operation. Offshore wind farms (OWFs), with their large capacity and fast controllers, have potential as innovative black-start units, thus, the need for a new design for OWFs. In this paper, challenges and possible solutions in integrating black start services into offshore wind farms will be presented. The first challenge is represented by the implementation of a self-start unit. The self-start unit should be capable of forming the wind farm power island and withstanding transient phenomena due to the equipment energisation. The investigated solution comprises grid-forming (GFM) converters in the wind farm design, which could be battery energy storage systems (BESSs) to also increase the service availability. The challenges are analysed using simulations on a wind farm, and the proposed solutions are discussed. It can be concluded that a hybrid system comprised of a BESS and an OWF, in combination with novel technologies such as GFM control, soft-charging, etc., represents a good proposal to provide black start services by OWFs.

## LIST OF ABBREVIATIONS

| | |
|---|---|
| BESS | Battery energy storage system |
| GFL | Grid-following |
| GFM | Grid-forming |
| HVDC | High-voltage direct current |
| MMC | Modular multilevel converter |
| OWF | Offshore wind farm |
| SRF | Synchronous reference frame |
| TSO | Transmission system operator |
| UK | United Kingdom |
| VSM | Virtual synchronous machine |
| WT | Wind turbine |

## I. INTRODUCTION

Power systems are currently experiencing a transition towards the decarbonisation of electricity generation through the large-scale deployment of renewable energy sources. Industries are facing a major re-thinking to cut emissions faster since global warming has to be limited to an increase of 1.5°C by reaching net-zero emissions in 2050, as per the Paris Agreement. Therefore, conventional power plants which run on fossil fuels are more and more often taken out of operation to accommodate higher shares of renewable energy generation. However, this weakens the resiliency of the power system after a blackout, as there are no black start sources ready to restore the grid in this scenario. On the other hand, it is not convenient to maintain conventional power plant systems on stand-by only for the possible risk of a blackout [1]. Consequently, economic considerations also point towards the integration of new types of black start service providers. Integrating black start services into non-conventional technologies to participate in the black start market could introduce a new revenue stream to operators and lower costs for electricity consumers, due to increased competition [1]. Therefore, the integration of a large volume of renewable energy sources as black start providers offers new opportunities in power system restoration planning.

High attention has recently been given to the black start, as it represents a crucial counteraction to the emergency state introduced by a blackout [2]. The occurrence of a massive power outage that includes the complete loss of generation, load and the transmission network serving the system loads, requires the use of selected generating stations with self-starting capability, i.e., black start sources, to get the system back into operation [3]. Power system blackouts are rare, but depending on how fast the power is restored, blackouts may have a great impact on the economy and society. Hence, a fast and reliable restoration scheme is a major concern to minimise economic and societal losses. Important organisations such as ENTSO-E [4], CIGRÉ [5] as well as the EU Commission [6] have addressed this concern.

Among renewable energy sources, wind energy is one of the fastest growing, thanks to its abundance and sustainability [7]. In particular, offshore wind farms (OWFs) represent an attractive solution at the global [7], European [8] and both Danish [9] and British levels [10]. In this context, OWFs, with their large capacity and fast controllers, have potential as innovative renewable-based black start units. Since the restoration time reduces exponentially with the availability of black start units, integrating black start capability into OWFs could significantly reduce the extent, intensity and duration, and thus the overall impact of blackout events [1]. Therefore, OWFs must gain the ability to participate in the restoration strategy as black start units. Furthermore, the black start is a crucial and extra service that is agreed upon separately from ancillary services. Thus, it is additionally paid, and thereby potentially increases the revenue for operators of OWFs.

The practical feasibility of successfully providing black start capability from an OWF still has to be proven, as no OWF is currently able to restore a black grid. Therefore, more research needs to be conducted. Some literature around the main topic can be found in [11, 12, 13, 14], where some discussions are made around OWF system configuration and control for a black start. However, no research is found, which

This work is part of an Industrial PhD project partially funded by Innovation Fund Denmark, project 9065-00238.

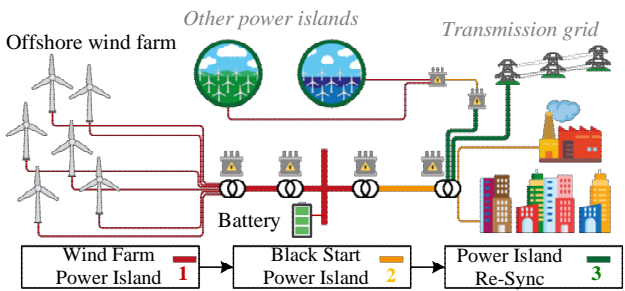

Fig. 1: Stages of a black start procedure implemented by an offshore wind farm equipped with battery.

discusses the challenges that black start functionality introduces in OWFs and presents possible solutions which could be taken into consideration and implemented. Nevertheless, innovative power electronic solutions, in combination with modern energy storage systems can enable this [15]. The increase and diversification of novel power-electronic-based devices, such as wind turbines (WTs), high-voltage direct-current (HVDC) transmission, STATCOMs and battery energy storage systems (BESSs) in particular, creates many opportunities for extended functionalities and innovation in power systems. An attractive novel technology is represented by grid-forming (GFM) convert control, which may be a viable option for the self-start of a power-electronic-based power system. In [16], the solution to apply GFM converters in OWFs is presented and lists the advantages of introducing a GFM unit onshore to reduce the costs and downtime of the OWF related to offshore equipment. Nevertheless, the GFM unit introduced onshore for ease of maintenance is a STATCOM, which does not have the advantages of additional energy storage as a large BESS has, which is required for system restoration. In [17], the idea of introducing a BESS in the wind farm design is presented to improve the performance of the grid-connected system. However, the battery is not a GFM unit and there are no discussions on black start operation. In , a novel type of device, consisting of integrated STATCOM and BESS is introduced in the OWF to perform island operation and black start. The device, defined as IBESS, is a GFM unit, nonetheless, no discussion on the additional solutions is shown to perform a real black start.

The rest of this paper (which is an extension of [20]) is organised as follows; an analysis of black start functionality in OWFs and transmission systems is given in Section II, where different stages are identified in the implementation of black start procedure by OWFs are outlined. Then, the selected challenges and possible solutions for OWFs as black start service providers are presented in Section III. These are classified and justified according to their different aspects. Black start simulation analysis on an OWF system inspired by and adapted to large projects in the UK like Hornsea I and II with integrated GFM battery and discussions are presented in Section IV. Discussion is made in Section V. Finally, concluding remarks are given in Section VI.

## II. BLACK START FUNCTIONALITY ANALYSIS

A black start comprises of different stages, going from a state of no power to a state of restored normal operation for the power system. The OWF operating as a black start unit needs to meet a certain set of requirements, referred to as black start requirements. As shown in [11], these requirements do not apply to OWFs currently as they are not conventional power plants. Nevertheless, an extended set of requirements has been proposed by some transmission system operators (TSOs), e.g., in the United Kingdom (UK) which are considered here [22]. When using an OWF as a black start source, the first stage represents the energisation of the wind farm without being connected to the main grid, which is shut down, i.e., working in island mode. This stage is referred to as wind farm power island. Once the island system is energised and stable, the energisation of the onshore transmission grid and block loads can start and thus form a black start power island. This is also a challenging operation, as the transmission grid itself consists of many transmission lines and large loads, such as substations, which have to be energised. After the energisation of a part of the system where the OWF is located, the synchronisation with other external power islands has to take place until the full system is synchronised and restored. These stages are shown in Figure 1. As they have different characteristics, these will be analysed separately.

### A. Wind Farm Power Island

After a total shutdown, the electrical circuits of the OWF are found in a state of no power, with all WTs disconnected. The first operation happening instructs the OWF operator that a black start operation needs to take place. There will be a time limit to connect to the onshore grid, e.g., two hours for the UK [22], within which the black start unit needs to be ready to energise the onshore transmission grid and block loads. The OWF has to have a unit able to self-start, which can deliver the first energisation power. This self-starter will have to energise first the OWF passive system, i.e., transformers, export and array cables, substations and other parts of the system that are needed for the black start. Afterwards, the energisation and synchronisation of the WTs can take place, as seen in Fig. 2. This procedure can be challenging due to the different devices present, i.e., reactive components, power electronic converters, and the switching operations involved. The switching of the reactive components can trigger transient phenomena such as inrush currents, which are problematic for power-electronic-based generation sources in case they exceed established limits in voltage and/or current especially.

Initially, the wind power island is a dead system, and therefore, the location of the self-starter, as well as the energisation strategy, are fundamental for a resilient black start strategy. Once energised by the self-start unit, the OWF is working as a wind farm power island, which is a very weak grid [18]. Once the system is stable and ready, the actual black start service seen from the grid can take place.

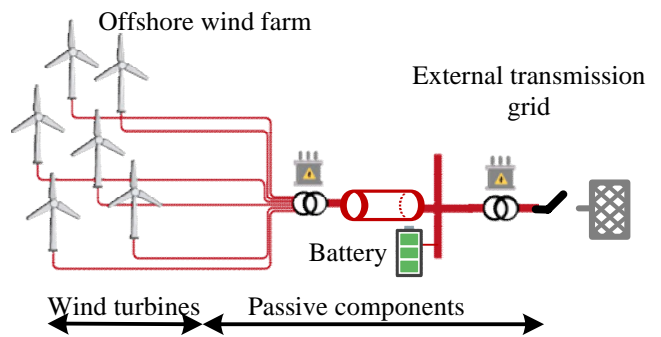

Fig. 2: Wind farm power island (Stage 1) of the proposed black start procedure split into two substages.

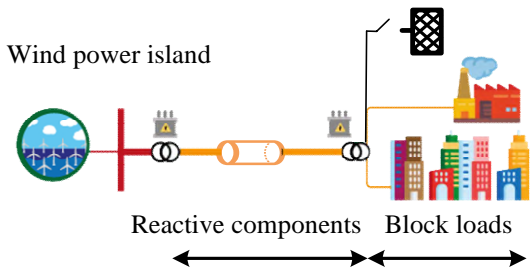

Fig. 3: Black start power island (Stage 2) of the proposed black start procedure split into two substages.

*B. Black Start Power Island*

This stage is the first where an actual system restoration is occurring as seen from the grid. At this stage, the whole OWF acting as a black start source has to energise both passive components of the transmission and distribution network, as well as block loads. This is schematised in Fig. 3. There are specific requirements for both, and they are different according to the country. In the UK, the reactive capability of the black start unit needs to be at least 100 Mvar exported from the OWF, and the block loading capability is set at a minimum of 20 MW. Due to the variability of the transmission network configuration, the OWF as a black start unit needs to be prepared for different scenarios and flexible in accommodating them. The challenge to solve at this stage consists mainly in the performance of the OWF, and all its power-electronic based sources, during the energisation of the system where phenomena like high inrush currents, transient overvoltages, resonances and loss of synchronism could jeopardise the black start power island stability. Once part of the onshore network is restored the OWF is working as a black start power island where the OWF is exporting power to the transmission grid while controlling the frequency of the island. The actual number of block loads will depend on the TSO requested service and the OWF power available at the instant of block loading. Furthermore, the considerations need to account for the OWF power availability for a minimum of three days, according to the TSO's needs in the overall system restoration.

After this stage, the interconnection with other power islands will occur to complete the restoration of the network.

*C. Power Island Re-Sync*

Initially, the OWF is controlling power consumption and frequency of the black start power island. When the TSO is ready for the energised islands to re-synchronise into one synchronous grid, the OWF operator has to initiate the grid synchronisation processes, as shown in Fig. 4. The transition from islanded to grid connection needs to be seamless, therefore the controller settings need to easily pass from islanded to grid-synchronisation.

The overall process in shown in the flowchart in Fig. 5.

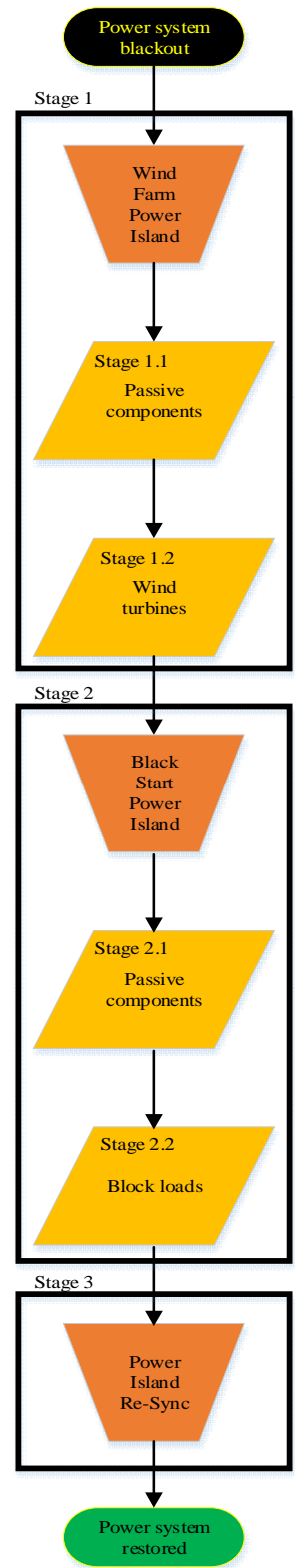

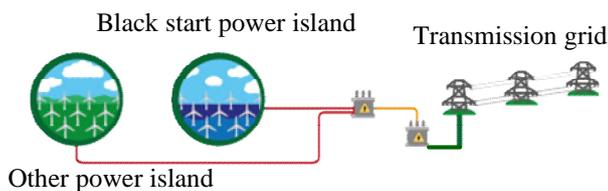

Fig. 4: Power island re-sync (Stage 3) of the proposed black start procedure.

Fig. 5: Flowchart depicting the main steps in the black start procedure by an offshore wind farm.

## III. CHALLENGES AND POSSIBLE SOLUTIONS IN INTEGRATING BLACK START INTO OFFSHORE WIND FARMS

The main challenge posed by the integration of black start capabilities into OWFs is the fact that OWFs are usually 100% power-electronic-based systems whose source of power is variable, which is in contrast to the conventional black start sources that are mainly synchronous generators.

Therefore, different challenges arise as presented in [11]. These challenges and their proposed solutions will now be analysed and discussed together with solutions.

### D. Self-Start Capability

Firstly, power electronic converters present in usual OWF designs are grid-following (GFL) converters, which do not have self-starting capability. This is because normally OWFs do not contribute to restoration strategies, and they are typically energised by the transmission grid. Hence, in order to integrate black start capabilities into OWFs, there is a fundamental need to introduce a device with self-start capabilities.

There are different alternatives and configurations that are possible [11]. The use of a new-generation power electronic converters with GFM control is highly interesting. These converters in the OWF could be storage units or GFM WTs. Another option is represented by the use of a conventional unit as a synchronous generator, which could be fuelled by diesel or, as more recently proposed, by biomass [1], or even green hydrogen. Nevertheless, more and more research is focused on the application of novel GFM control strategies to black start, instead of synchronous generators. This current trend is mainly due to the fact that many industries are trying to support sustainable goals, by implementing carbon neutral solutions in their operations and supply chain, and therefore want to avoid the use of fossil fuels, such as diesel, as much as possible. Furthermore, biomass-powered synchronous generators are typically available on a much smaller scale than conventional units [1], thus, this may affect their capability. Therefore, the use of GFM converters remains the main focus.

### E. Service Availability

Secondly, the availability problem is of high concern. As the wind is the source of power in an OWF, the black start service availability highly depends on wind variability. For the UK, the first proposal for the black start availability requirement is 90%, which is the same established for conventional black start sources. As presented in [26], the average amount of procured wind capacity to secure a sustained 500 MW of power for 24 hours for the $90^{th}$ percentile is higher than 10 GW. This is certainly an extreme amount, and it is also stated that a very large diversity in available power could be found also from two wind farms with very similar total capacity, depending on their geographical location. Nevertheless, the challenge posed by the service availability requirement is understood, although it may be that TSOs would benefit from modifying the requirements to find the optimum balance between their specific needs and ensure that enough OWFs can contribute with a black start. This would surely also need to be examined thoroughly by the OWF system operator. Combining energy storage with OWFs could be a potentially attractive way to boost the contribution that OWFs could make. Generally, the integration of energy storage can contribute to a more efficient and stable power system. The choice of adequate energy storage is based on various parameters, such as accurate technical and economic evaluation of deployed energy storage and geographical considerations. In [19], an overview of all existing energy storage systems is carried out. Out of various types of energy storage systems, BESSs have recently been considered the most suitable technology due to numerous advantages. Their scalability, response time and ability to absorb and deliver power to the system is making them a good option for smoothening wind power output and increasing power system stability. Moreover, the cost of this type of technology has been constantly decreasing and it is expected that the trend will continue in the future.

Lately, numerous projects involving BESS technology are announced, are under construction or are ongoing worldwide. The most recent one involved the installation of 150 MW/189 MWh Li-ion BESS called Hornsdale Power Reserve, co-located with the Hornsdale Wind Farm. It is considered the largest Li-ion BESS in the world and its purpose is to provide power system stability services in South Australia. Moreover, the Dalrymple BESS, also known as the ESCRI-SA project, is a 30 MVA/8 MWh GFM BESS with microgrid automation, controlled with the virtual synchronous machine (VSM) algorithm. This BESS can store excess generation from Wattle Point Wind Farm to charge and perform the necessary services.

As presented in [20] by the IBESS project, the combination of OWF and integrated BESS and STATCOM can deliver a black start with a high level of availability. Nevertheless, this depends on the OWF capacity and availability level required by the TSO, as discussed above. Analytical results show that a partially energised OWF in combination with BESS is able to provide black start with a level of availability higher than 80%. Meanwhile, availability higher than 90% can be provided with a larger BESS and fully energised OWF, which is costly and depends on wind conditions [21]. As a consequence, it can be understood that adequate competition in the black start market to enable renewable energy sources to participate in this new service depends on the selected technical requirements. A lower level of availability required by the TSO allows more OWF generators to contribute to black start and ensure more competition.

### F. Inertia Provision

Inertia provision has been added as a black start requirement. This is challenging for conventional OWFs as they are 100% power electronic-based systems. Nevertheless, this requirement can be fulfilled either by real or virtual inertia. Virtual inertia, the requirement can be achieved by implementing inertia emulation in the converter control. More specifically, the WTs and/or BESS in the OWF can be controlled as VSM and deliver inertia emulation when needed. Both GFL and GFM control can emulate inertia and examples are described in. Practical applications of VSM GFM in power electronics can be found for BESSs in and WTs in.

Furthermore, VSM algorithms have been proposed for STATCOMs to emulate inertia. In this context, STATCOMs could behave as synchronous condensers without the drawback of active power losses. Innovative types of STATCOMs can support the system with more than just dynamic voltage regulation. One example is the SVC PLUS® with power intensive energy storage, which is a combined STATCOM with supercapacitors which is able to provide both voltage and frequency support [38]. These capabilities could be exploited and used for inertia emulation as well.

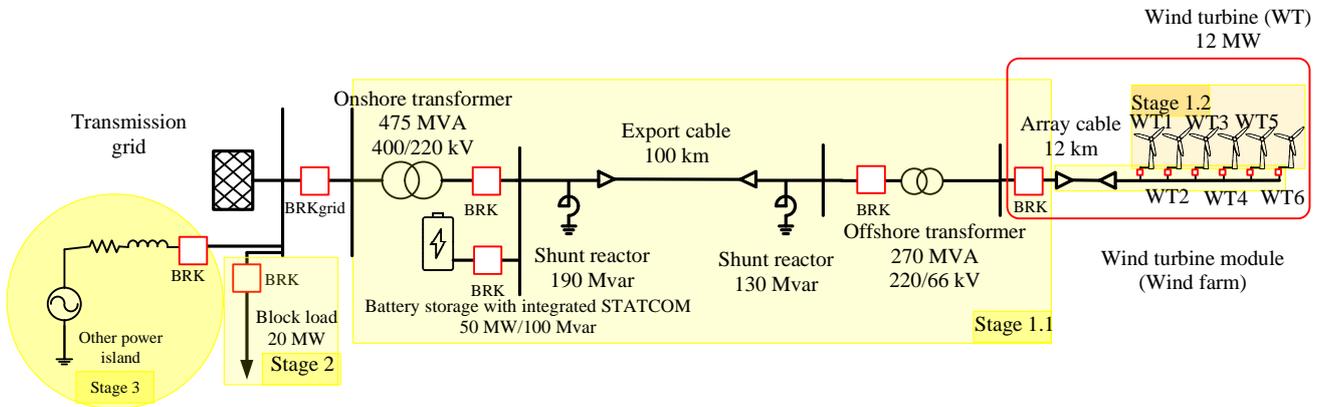

Fig. 6: Single-line diagram of the offshore wind farm and battery system used for the black start studies.

*G. Control, Stability and Interoperability*

With a whole new service to provide, the control and interoperability of the OWF system also adds a challenge. Depending on the numbers of self-starters, there will be a need to coordinate the interoperability, both during the energisation and the operation of the system. In particular, the challenge gets tougher as the islanded OWF system represents a very weak grid, compared to the onshore transmission grid, which is usually the reference for the power electronic control. Therefore, issues like energisation sequence and island frequency control are of main importance. In this regard, the possibility to combine a GFM BESS together with GFM WTs has the potential to ensure stable operation also in weak grid conditions. The GFM unit(s) will have firstly to coordinate the response, which could be primarily assigned to the BESS, in order to avoid relying on wind conditions, and then be shared with the WTs. During operation, the interoperability could be guaranteed keeping the system coordinated by a high-level control, which applies communication, or by droop strategies for example. Moreover, as the system parts get energised, the need for an adaptive control may emerge in between these settings. Hence, the complexity of the proposed solution is increased. After deployment of a black start demonstration to show that the implemented solutions are feasible, also automation and communication settings need a viable solution to have a faster black start procedure with reduced error in the case of a real blackout scenario.

BESSs are considered as one of the prominent solutions for stability maintenance in a renewable energy system. Thus, their use is expected to be relevant in OWF black start.

There are many different types of GFM control strategies that have been presented in the state-of-the-art literature and that could be applied for OWF black start. In between these strategies, the VSM type of GFM control has achieved successful application, e.g., in field tests for GFM WTs and the Dalrymple BESS working in combination with Wattle Point Wind Farm in South Australia.

*H. Transient Behaviour*

Transient phenomena are involved in every energisation operation related to OWF black start. Transients due to the switching operations can be challenging for the system to survive. Transient overvoltages can lead to equipment failure or damage that may hinder the successful implementation of the restoration plan. Energising equipment during black start conditions can result in higher overvoltages than during times of normal operation [3]. These transient overvoltages originate from energising operations and equipment non-linearity, e.g., transformer saturation. The first challenge is represented by the wind farm power island energisation. To contain the energisation transients, novel energisation methods like soft charging can be applied with GFM units. After that, another challenge is the energisation of the transmission system and block loads. If the load has been de-energised for several hours, the inrush current upon re-energising the load can be as high as eight to ten times normal [3]. Hence, a detailed black start strategy and simulation plan will have to confirm that the OWF system will be able to cope with a black start.

*I. Harmonic Performance*

The combination of different power converters, long cables and other passive elements also introduces issues for the harmonic performance of the system. This is because power system resonances may be excited during the restoration procedure, potentially leading the system to instability conditions. Therefore, the analysis of the black start strategy needs to consider this challenge to mitigate the possible effects of harmonic distortion.

The main solution to this challenge can be active harmonic filtering, which is adaptive to topology changes and system loadings. This represents a flexible harmonic mitigation measure and can be successfully applied to the OWF during black start.

IV. BLACK START STUDIES

In a real black start scenario where the black start source is an OWF with a BESS as the main self-start unit, a general strategy needs to be implemented. As the integration of BESS together with WTs represents a hybrid generation system, the main role of the master could be given to the BESS. In this way, the BESS is the first element to self-start once the command of black start is instructed by the TSO, and energises the OWF, forming the wind farm power island. According to this procedure, the BESS supplies the initial power, while the WTs will then charge the BESS to maintain an optimal state of charge. Forming the black start power island, i.e., energising 20 MW block loads, the BESS will also be the one providing the initial energy, which will then be passed to the WTs. This procedure can ensure high resiliency as the main power supply relies on one single element, where the necessary energy storage is controlled, and then it is backed up by a large OWF. Some of these presented challenges can be illustrated by a simplified OWF system. Time-domain simulations of the black start system have been developed in PSCAD/EMTDC.

*J. Offshore Wind Farm Model Description*

The studied system is shown in Fig. 4. The OWF model is simplified and scaled down from the Hornsea Projects in the UK, where the large wind farm is very far from shore, i.e., over one hundred kilometres from shore.

In the model, the export system is comprised of 220-kV cables, for a total length of 100 km. The first section of the export cable is a 40-km long land cable, which is considered as a three single-core cable laid Al XLPE 1200 mm$^2$ cable in flat formation. The remaining 60-km long cable is a three-core submarine Cu XLPE 1600 mm$^2$ cable, represented as single-core cables and piped in three-foil. The export system is represented with frequency-dependent models, as it is useful for studies wherever the transient or harmonic behaviour of the cable is important.

Due to the large capacitive behaviour of cables, shunt reactive power compensation equipment has also been considered. There is a shunt reactor located onshore sized at 190 Mvar, and one offshore sized 130 Mvar, both at 220-kV level.

The offshore end of the export cable is connected to an offshore 220-kV busbar, which in turn is connected to a 220/66 kV, 240 MVA transformer. Only one WT module is considered for simplification. This WT module is comprised of one array, which is connected to the transformers via a 66-kV busbar. Each array consists of 6 WTs, for a total of 18 WTs in the model. The WT model has been taken from [41] and adapted to the studies. This model represents a 12 MW WT operating as GFL unit (as they normally do nowadays). As a simplification, WTs are modelled by their grid-side converter model and relative controls, together with their filters and transformers. The WT control is implemented in the synchronous reference frame (SRF), i.e., *dq* frame. The WTs operate in voltage regulation mode, that means defining the reactive current reference proportionally to the voltage deviation from the nominal value, which is 1 pu at the 66-kV transformer bus. A droop frequency regulation of 5% in steady state is implemented.

To speed up the simulation time, only one array has been implemented with six separated WTs. Array cables connect the WTs to the 66-kV side of the offshore transformer. These have been assumed to be 12-km long cables, and have been modelled as T models, to reduce the computational burden of the simulation.

At the onshore end of the cable the 400/220 kV, 475 MVA onshore transformer is connected. This connects the OWF to the onshore grid, which is only considering a 20 MW block load for simplification. In this way, it is possible to first simulate the wind farm power island and subsequently the black start power island and its impact on the self-starter, intending to consider a BESS for this application. The saturation characteristics are considered for the 400/220 kV and the 220/66 kV transformers.

*K. Grid-Forming Battery Model Description*

As stated, a GFM BESS is implemented in the OWF to integrate black start capabilities into the system. In this scenario, the GFM BESS is a large unit connected onshore at the 220-kV bus. This is mainly since usually the offshore substation of an OWF does not allow for large equipment such as a BESS. The size could be lowered to make it feasible to install offshore, however, the size of a BESS possible to install offshore would probably be too small for an effective black start procedure. Furthermore, the resiliency of the system is higher when having the BESS onshore, as the maintenance and reparations can be faster due to easy access to the location.

The GFM BESS is modelled as the average model of their grid-side converter and controller. The unit is imagined to be a large BESS unit, thus using a converter based on modular multilevel converter (MMC) technology. This directly connects to the 33-kV level and only needs one 33/220-kV step-up transformer to be connected to the 220-kV bus. Furthermore, thanks to the MMC technology, only an L filter is needed at the converter output to mitigate it. The model is inspired by [20], and consists of a 112 MVA unit, where the active power nominal rating is 50 MW, while the reactive power nominal rating is 100 Mvar, to allow for dynamic var regulation. Its controller is shown in Fig. 7, where it is possible to see that the chosen GFM algorithm sets the voltage magnitude v$_{abc}$ and phase angle θ used in the SRF transformation. The former is made of the reference voltage, which is compared to the RMS value of the voltage at the onshore 220-kV bus. This is multiplied by the gain k$_V$, filtered and then fed to the reference voltage again which results in the direct voltage v$_d$, while the quadrature voltage v$_q$ is set and kept constant to 0. The second part of this GFM converter is made of the power/frequency P/ω droop. This simple controller is used as a preliminary example to show some meaningful results. On the other hand, its simplicity has many limitations, as it does not provide additional inertia and damping, that may be required for a smooth operation, as well as current or voltage limitations, which on a real power electronic device will damage the semiconductor switches.

All the main parameters of the model implemented for these studies are shown in Table I.

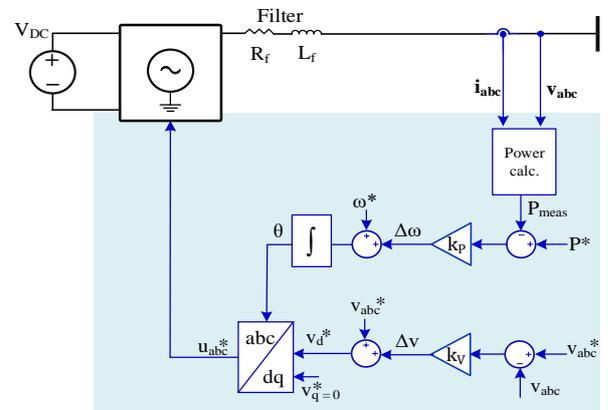

Fig. 7: Controller of the grid-forming battery used in this study.

TABLE I. CONVERTER PARAMETERS IMPLEMENTED IN THE STUDIED MODEL.

| Parameter Description | Value |
|---|---|
| **Battery Energy Storage System (BESS)** | |
| Nominal apparent power | 112 MVA |
| Nominal active power | 50 MW |
| Nominal reactive power | 100 Mvar |
| Nominal RMS voltage | 33 kV (line-to-line) |
| Filter inductance | 0.1 pu |
| Filter resistance | 0.025 pu |
| Transformer resistance | 0.004 pu |
| Transformer inductance | 0.1 |
| Nominal angular frequency $\omega_0$ | $2\cdot\pi\cdot50$ rad/s |
| Voltage controller gain $k_V$ | 10 |
| Power controller droop gain $k_P$ | 5% |
| **Wind Turbine (WT)** | |
| Nominal active power | 12 MW |
| Power factor | 0.9 |
| Filter inductance | 0.1 pu |
| Filter resistance | 0.005 pu |
| Filter shunt resistance | 0.003 pu |
| Filter capacitance | 0.075 pu |
| Transformer resistance | 0.0054 pu |
| Transformer inductance | 0.1 pu |
| PLL proportional gain | 0.1 |
| PLL integral gain | 2 s |
| PLL filter | 500 rad/s |
| Nominal angular frequency $\omega_0$ | $2\cdot\pi\cdot50$ rad/s |
| Current controller proportional gain | 0.2 |
| Current controller integral gain | 5 s |
| Voltage controller droop gain | 0.05 |
| Voltage controller proportional gain | 0.3 |
| Voltage controller integral gain | 0.2 s |
| AC voltage controller proportional gain | $10^{-4}$ |

*L. Simulation Description*

In this study, the black start operation of the OWF, starting from the GFM BESS located onshore will be shown. The main objective is to demonstrate some of the challenges presented in § III and discuss the related possible solutions. The black start strategy starts from the system completely shut down, i.e., de-energised. The BESS is the self-start unit and provides power to the rest of the system, comprising the WTs. As discussed, the first step for this black start operation is to form the wind power island. One of the many challenges with energisation is to deal with the transient behaviour, as stated. Therefore, the application of the method of soft charging is presented. Furthermore, the energisation of the array cable and related WTs is presented, as it is not possible to soft charge due to the power electronic nature of the WTs. This part of the black start procedure represents the wind farm power island.

The energisation procedure is schematised in Table II. During soft charge, the whole passive islanded network is energised, from the onshore transformer to the offshore transformer with all the breakers closed and the BESS forming the grid voltage from 0 to 1 pu in 0.5 s.

The WTs are connected with an interval of 3 s in order to reduce the simulation time, however, it would be longer time in reality. It would take longer in real life due to the physical time constraints involved, especially by the mechanical part, in the range of tens of minutes. All the WTs are initialised having as DC-link voltage reference 1 pu and the voltage at the 66-kV bus at 1 pu, while the active power reference is left equal to 0 pu for the whole simulation to have the BESS as main power unit. After the passive network and WTs are energised, a 20 MW block load connected at 400 kV is energised. This represents the start of the black start power island, when the wind farm energises parts of the transmission grid. The simulation is run for 25 s. The simulation is run for 25 s. The energisation procedure is schematised in Table II.

TABLE II. ENERGISATION PROCEDURE WITH SWITCHING INSTANTS USED IN SIMULATIONS.

| Event | Instant | Stage |
|---|---|---|
| Soft charge of the system from 0 to 1 pu voltage | From 0 s to 0.5 s | Wind farm power island |
| Energisation of WT1 | 1 s | Wind farm power island |
| Energisation of WT2 | 4 s | Wind farm power island |
| Energisation of WT3 | 7 s | Wind farm power island |
| Energisation of WT4 and WT5 | 10 s | Wind farm power island |
| Energisation of WT6 | 16 s | Wind farm power island |
| Block load energisation | 19 s | Black start power island |

*M. Simulation Results*

First and foremost, the BESS energisation of the export cable and shunt reactor system by means of hard switching is shown. In Fig. 8.a, the BESS voltage is formed and at 1 s the switching operation takes place, starting a long transient characterised by voltage exceeding the ±10% threshold and having unbalanced phases due to the sudden connection of the long export cable, which represents a capacitive component, together with the inductive compensation of the shunt reactors. Furthermore, Figure 8.b shows the current transient which is even more severe and will violate the current capabilities of the BESS converters, as it reaches over a.

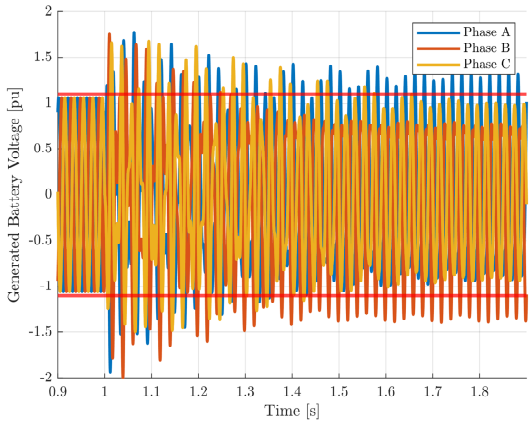

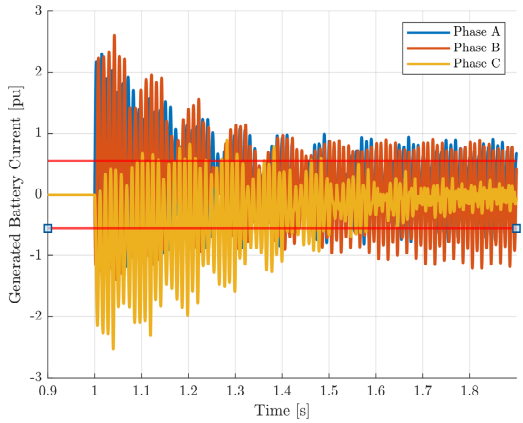

b.

Fig. 8: Simulation results from the grid-forming battery energising the export cable and shunt reactors by hard switching measured at the battery point of connection: a) Voltage from the grid-forming battery, b) Current from the grid-forming battery.

2 pu which is not acceptable for power electronic components.

As a comparison, the simulation illustrated in Table II is presented. In the simulations, firstly the BESS energises the passive network via soft charge. In Figure 9.a, the ramp-up of the voltage in 0.5s is shown. The controller has a small overshoot in the first 0.1 s, which is also reflected in the voltage and the current waveforms, while subsequently follows the reference, settling at 0.994 pu with small oscillatory behaviours, an aspect that could be improved for a real black start implementation. The voltage and current during this process are measured at the BESS point of connection and are shown in Figures 9.b and .c, respectively. While the voltage is increasing smoothly, the currents have some oscillations, indicating that the system needs more damping. This energises the export cable system and the offshore grid. The BESS also provides the active and reactive power for the system, which are shown in Figures 9.d and e., respectively. During the energisation, the BESS supplies the maximum active power of 6.1 MW, while it absorbs the maximum reactive power of 11.5 Mvar, which is generated by the high voltage export cables. This is the excess capacitive power considering that in no load no series inductive power appears, therefore the capacitive reactive power is in excesses. Since the BESS is controlled to regulate the HV terminals, the BESS is absorbing 47.5 Mvar reactive power to decrease the voltage. Thus, for energising the offshore grid, the active power requirement is very small, but it needs to absorb also a large amount of reactive power. This indicates that in a real black start scenario, the passive var compensation needed should be kept to a maximum (e.g., by using suitably adjustable reactors rather than fixed reactors) for the BESS to supply/absorb the minimum amount of reactive power as possible. This will help in maintaining the BESS with a lower load during the whole operation. In the steady state, at 1.0 s, the BESS supplies 2.5 MW and absorbs 11.5 Mvar at 50 Hz, as seen in Figure 11.d and .e.

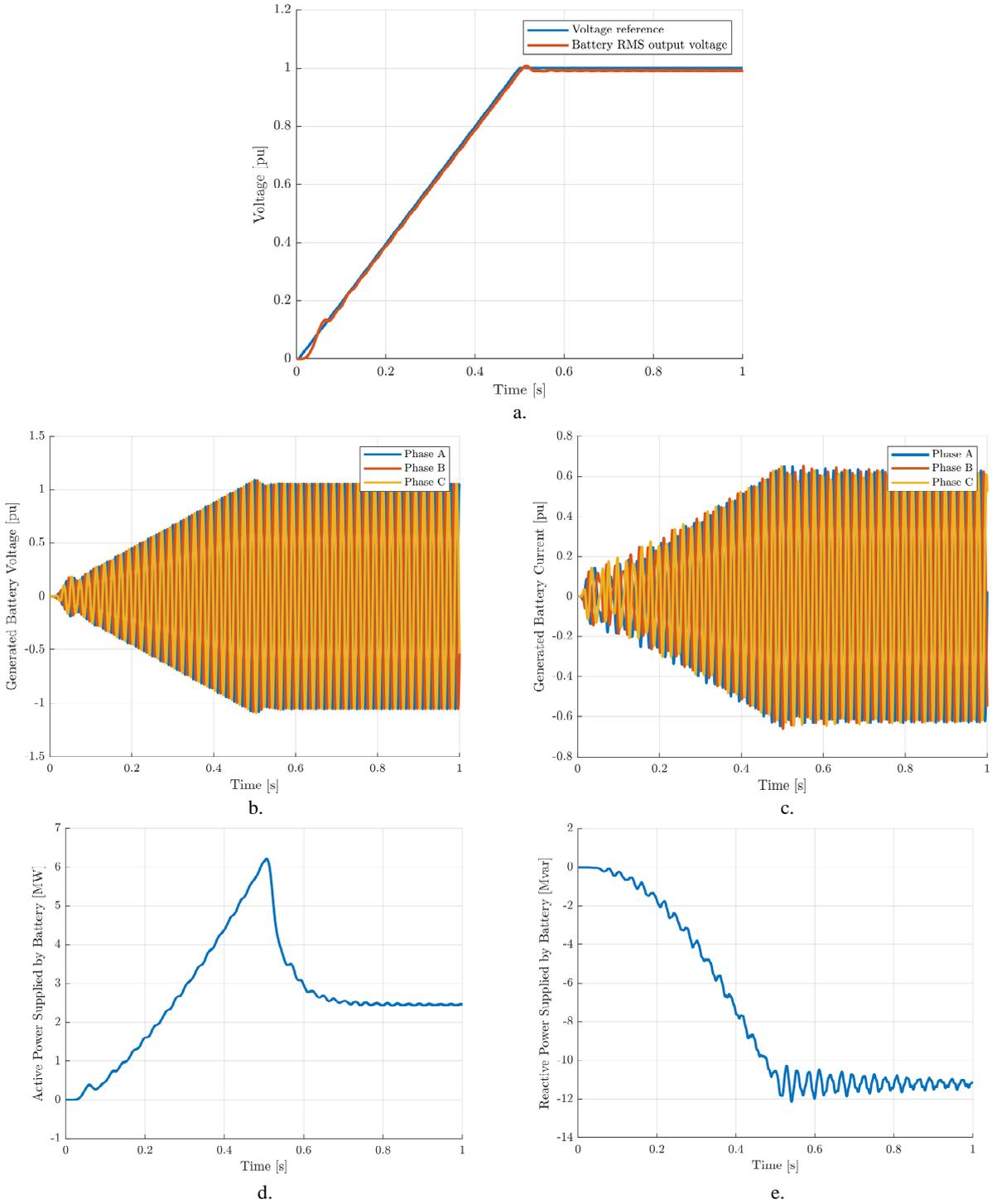

Fig. 9: Simulation results from the soft charge of the wind farm passive system in Figure 2 by the grid-forming battery measured at its point of connection: a) Voltage from the grid-forming controller, b) Generated battery voltage, c) Generated battery current, d) Active power supplied by the battery and e) reactive power supplied by the battery.

After the soft charge, the initialisation and synchronisation of the WTs take place. The process is synthesised in Table II and the resulting BESS voltage and current are shown in Fig. 10. In Fig. 10.b, the voltage waveform is shown. It can be seen that the more WTs are connected the more oscillatory behaviour on the BESS is seen. This arises from the more power electronic unit working in parallel in this islanded network. This is a point that could also be solved by extra damping and inertia in the control system. The first WT, WT1, is synchronised at 1 s. Consequently, there is a slight rise in the voltage of 0.1 pu. The requirements specify also voltage and frequency control to be respected during black start, and for example for the UK, the voltage requirements relate to ±10% (shown with red lines in the figures). Fig. 8.a shows also that the frequency set by the BESS has a small increase for every WT connected, reaching around 50.006 Hz for the last group. These deviations do not exceed the 47-52 Hz range given by NGESO. However, this could be kept at 50 Hz by a second level controller that regulates the power island. For every WT connected, also a large increase of the current takes place as seen in Fig. 8.d. Nevertheless, it does not exceed the limits of 10% (shown as well with red lines in the figures) as the BESS is working at almost no load. The transient disturbance becomes larger and larger for every WT that gets connected.

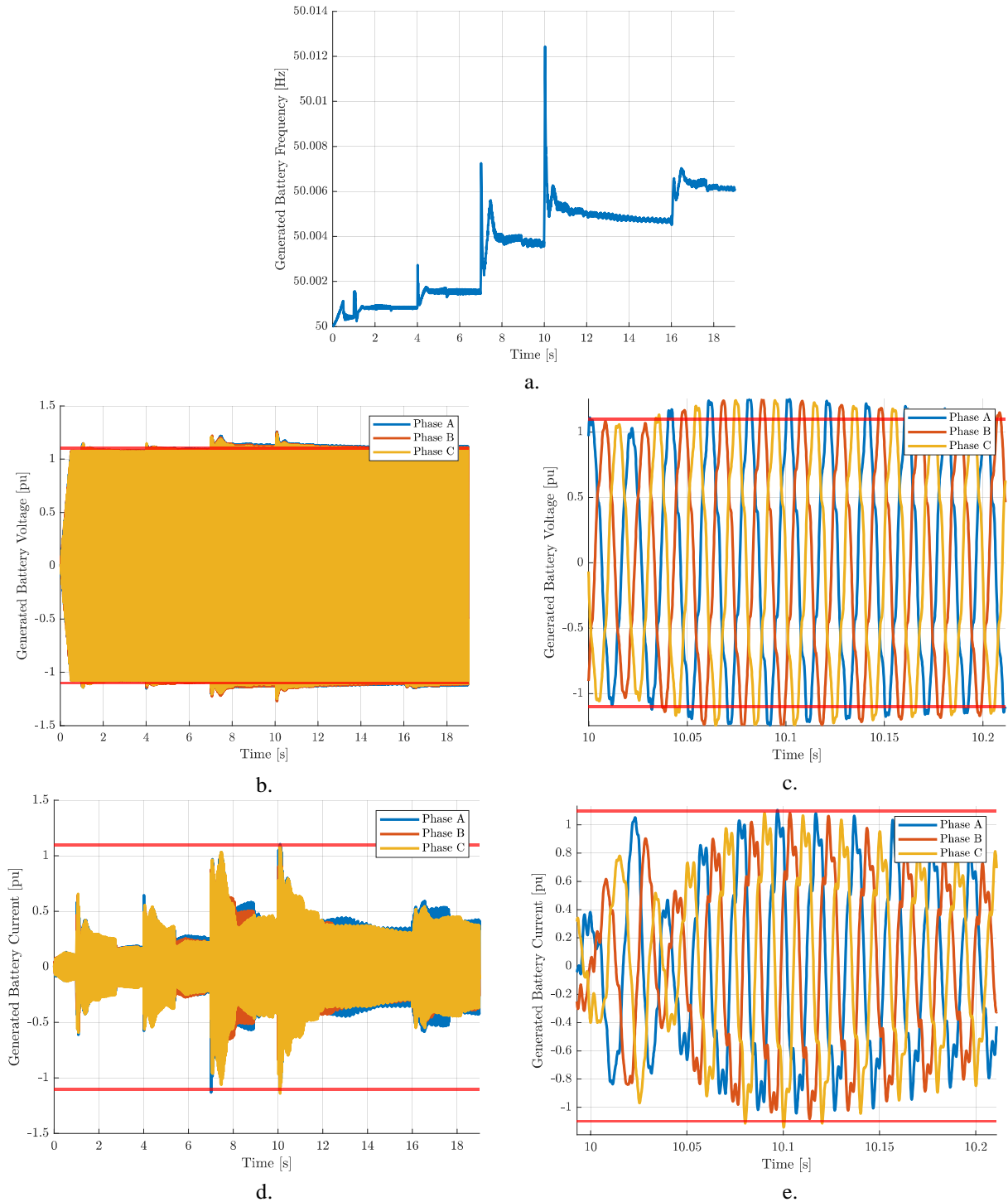

Fig. 10: Simulation results from the initialisation of the wind turbines (WTs) measured at the battery point of connection: a) Frequency formed by the battery, b) Voltage from the grid-forming battery while the WTs are initialised, c) Zoom of the voltage waveforms during the initialisation of WT3 and WT4, d) Current from the grid-forming battery, e) Zoom of the current waveforms during the initialisation of WT3 and WT4.

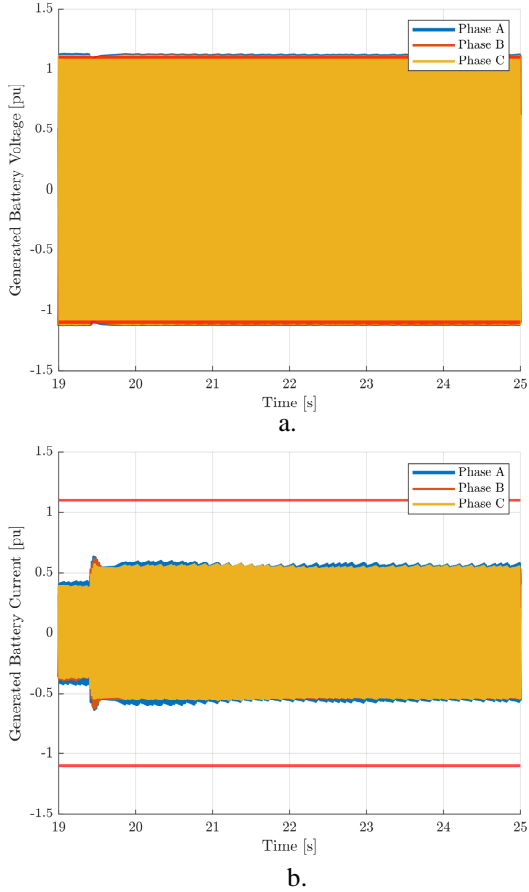

Fig. 11: Simulation results from the block loading measured at the battery point of connection: a) Voltage from the grid-forming battery, b) Current from the grid-forming battery.

It is shown in Figures 8.c and .e how the BESS voltage and current evolve when WT3 and 4 are connected at 10 s. It should be noted that the help of a converter control in the BESS with current and voltage limiter could avoid the impact on these transients of the power electronic hardware, which cannot tolerate overcurrents as more traditional synchronous generators. A further transient with the relevant increase of the current supplied by the BESS appears at the energisation of the 20-MW block load at 19 s, where the current goes from 0.46 pu to 0.57 pu in the steady state, as shown in Fig. 9.b. This is an important aspect to consider when implementing systems with BESSs as the current capabilities of this type of device are crucial during a black start scenario. The most important thing to consider is that the BESS does not exceed its rated current and collapses both in wind farm power island and black start power island mode. In these conditions, the system has formed a black start power island as the BESS and OWF are energised and supply block loads from the transmission system. The BESS power reference has been updated to 20 MW at 19 s and the controller has been able to follow this change in the system.

During the whole simulation, the BESS was found to be able to form the wind farm power island and the black start power island as the black start unit. The BESS can cope with the different events, i.e., WT initialisation and block loading, even though a simple controller is used. The limitations shown in the simulation are part of the challenges previously discussed, e.g., transient behaviour, low damping and inertia, exceeding voltage and current limits. These may be counteracted with a more sophisticated design of the BESS system with related GFM control. Thus, it can be seen that a BESS may be the way of integrating black start capabilities into OWFs.

## V. DISCUSSION

In this analysis, the black start procedure started by a GFM BESS implemented in an OWF is discussed. The main point has been to highlight the challenges that such a system may experience and discuss possible solutions, together with the analysis of a simulation case. As seen, such an islanded system has to rely on the GFM controller of the BESS to energise the whole passive grid and WTs. It has been shown that the operation can be challenging and that a relevant GFM controller has to be designed. In this simulation example, the system shows a performance that could be improved by the application of virtual inertia and damping in the control loop, i.e., as in VSM type of implementations. Furthermore, reactive power regulation and current can be a limitation. Therefore, the controller should have overcurrent limitation to prevent the BESS from damage.

With some adjustments, GFM converters can be a viable solution to integrate black start capabilities into OWFs, especially if applied on a device with a large capacity, like a BESS. From this analysis, it is observed that a dynamic procedure such as a black start has to be thoroughly investigated to make sure that the outcome has high resiliency and success rate. It is not currently implemented nor already proven that OWFs with a combination of BESS, STATCOM controlled as VSM constitute a resilient black start service provider. However, analytical simulations and hardware-in-the-loop types of demonstrations will be valuable solutions for proof of concept of the system before any actual implementation. To achieve a real black start by OWFs in the near future, Table III presents the main points to be addressed.

TABLE III. CHECK LIST OF MAIN COMPONENTS AVAILABLE, TESTED AND/OR THE REQUIRED FURTHER DEVELOPMENT.

| Component / Type of needed developments | Grid-forming battery | Grid-following wind turbines |
|---|---|---|
| Simulations | Additional performance such as virtual inertia and damping | Capability to work in island mode with very weak system |
| Lab test | Capability to energise passive components and large loads | Capability to work in island mode with very weak system |
| Field test | Complete black start testing procedure | |

## VI. CONCLUSION AND FUTURE WORK

This paper demonstrates the use of BESS in the islanded operation and black start of an OWF. Even though hard-switching cannot be supported by the BESS due to a large transient current of over 2.0 pu, soft-charging can be used to energise the whole OWF. As the BESS is set to regulate the RMS voltage at the high voltage terminals of its transformer, it absorbs all the excess reactive power generated by the export cables in the OWF. Whenever a new WT unit is switched in, it creates a momentary transient in the voltage which affects the voltage throughout the OWF. The BESS is fast enough to respond to all voltage events. All while the BESS supplies the active power losses. When the WT units start generating power, the BESS automatically absorbs the excess power. In response to the BESS power variation, the BESS frequency changes according to the GFM control. In turn, the WT units respond to frequency deviations as that is enabled. Thus, the grid forming control of the BESS is demonstrated to start and operate the islanded OWPP successfully. More research pointing towards this novel type of service in OWFs needs to be conducted. The possible solutions discussed are to be applied and thoroughly analysed. In this way, they can be considered valid and worth testing. Nowadays, different implementations may allow a successful black start provision by OWFs. Different solutions have been explored and proposed, also based on existing systems with similar purposes of black start applications.

The investigated system shows the challenges, which need to be overcome before implementing this system in real life. It is found particularly interesting to implement the discussed solutions, i.e., GFM control, VSM, soft-charging, active harmonic filtering, etc. As further work, the GFM control to be applied can also provide inertia and damping, as a type of VSM control. Furthermore, the application of GFM WTs, can be studied, giving the system higher flexibility in terms of interoperability of the black start system.

ACKNOWLEDGEMENT

The authors acknowledge the comments and inputs from M. Kazem Bakhshizadeh, Mikkel P. S. Gryning, Karsten H. Andersen and Philip Johnson, Ørsted.